\documentclass[a4paper,12pt]{article}
\DeclareMathSizes{12}{12.5}{10}{10}
\usepackage[left=2.5cm,bottom=3cm,right=2.5cm,top=3cm]{geometry} 

\usepackage{overpic,youngtab}
\usepackage{subfigure}
\usepackage[latin,english]{babel}
\usepackage{amsmath}
\usepackage{amssymb}
\usepackage{epstopdf}
\usepackage{graphics,psfrag,rotating}
\usepackage{mathabx}
\usepackage{graphicx}
\usepackage{dcolumn}
\usepackage{float}
\usepackage{pdflscape}
\usepackage{array}
\usepackage{booktabs}
\usepackage{amscd} 
\usepackage{mathtools}
\usepackage{fancybox}
\usepackage{fix-cm}
\usepackage[colorlinks=true,citecolor=blue,,linktocpage=true,linkcolor=blue,urlcolor=black]{hyperref}
\usepackage{tikz} 
\usetikzlibrary{matrix} 
\usetikzlibrary{positioning} 
\usetikzlibrary{arrows}
\usepackage{titlesec}
\usepackage{abstract}
\usepackage{centernot}

\def\be{\begin{equation}}
\def\ee{\end{equation}}
\def\bea{\begin{eqnarray}}
\def\eea{\end{eqnarray}}
\def\pd{\partial}
\def\a{\alpha}
\def\b{\beta}

\def\d{\delta}
\def\m{\mu}
\def\n{\nu}
\def\t{\tau}

\def\l{\lambda}

\def\r{\rho}

\def\vx{\vec{x}}
\def\bR{\bar{R}}

\def\bn{\bar{\nabla}}
\def\bR{\bar{R}}

\def\s{\sigma}
\def\e{\epsilon}
\def\bi{\begin{itemize}}
	\def\ei{\end{itemize}}
\def\bg{\bar{g}}

\definecolor{ogreen}{rgb}{0,0.7,0}

\DeclareGraphicsExtensions{.pdf,.png,.jpg,.gif,.jpeg}
\def\be{\begin{equation}}
\def\ee{\end{equation}}
\def\bea#1\eea{\begin{align}#1\end{align}}
\def\pd{\partial}
\def\a{\alpha}
\def\b{\beta}

\def\d{\delta}
\def\e{\epsilon}

\def\m{\mu}
\def\n{\nu}
\def\t{\tau}

\def\l{\lambda}

\def\r{\rho}

\def\vx{\vec{x}}
\def\va{\vec{a}}
\def\bR{\widebar{R}}

\def\bn{\widebar{\nabla}}
\def\bR{\widebar{R}}

\def\s{\sigma}
\def\e{\epsilon}
\def\t{\tau}
\def\bi{\begin{itemize}}
	\def\ei{\end{itemize}}
\def\bg{\widebar{g}}

\usepackage{authblk}
\usepackage{setspace}

\begin{document}
	
	\vspace*{-1cm}
\phantom{hep-ph/***} 
{\flushleft
	{{FTUAM-21-xx}}
	\hfill{{ IFT-UAM/CSIC-21-96}}}
\vskip 1.5cm
\begin{center}
	{\LARGE\bfseries  Some  consequences of the Kerr-Schild gauge.}\\[3mm]
	\vskip .3cm
	
\end{center}

\vskip 0.5  cm
\begin{center}
	{\large Enrique \'Alvarez and  Jes\'us Anero.}
	\\
	\vskip .7cm
	{
		Departamento de F\'isica Te\'orica and Instituto de F\'{\i}sica Te\'orica, 
		IFT-UAM/CSIC,\\
		Universidad Aut\'onoma de Madrid, Cantoblanco, 28049, Madrid, Spain\\
		\vskip .1cm

		\vskip .5cm
		
		\begin{minipage}[l]{.9\textwidth}
			\begin{center} 
				\textit{E-mail:} 
				\tt{enrique.alvarez@uam.es} and
				\tt{jesusanero@gmail.com} 
			\end{center}
		\end{minipage}
	}
\end{center}
\thispagestyle{empty}

\begin{abstract}
	\noindent
Generalized Kerr-Schild nilpotent deformations of an arbitrary background spacetime $\overline{g}_{\m\n}(x)$ are considered. Those are of the form $\overline{g}_{\m\n}+l_\m l_\n$ with $l^2=0$. The relationship between the Ricci tensor of the background metric and the Ricci tensor of the deformed metric is found exactly. It consists of two terms, one is essentially the Fierz-Pauli operator, and the other is new. When the background is flat, the Kerr-Schild family is recovered. Novel examples for more general backgrounds (even including some simple sources) are discussed.	
\end{abstract}

\newpage
\tableofcontents
\thispagestyle{empty}
\flushbottom

\newpage
\section{Introduction}
The importance of finding exact solutions of Einstein's equations of General Relativity (GR) cannot be overestimated. In spite of the portentous advances in numerical solutions, we still need them to examine global properties of the spacetime such as horizons, singularities, etc. They also serve as useful templates for simulations. Unfortunately exact solutions are few and far between. Any effort to find new such solutions is potentially useful in the endeavor of reaching a better understanding of the physics of gravitation. 

The vacuum equations of GR are just the vanishing of the Ricci tensor of the spacetime
\be
R_{\m\n}=0
\ee
which are nonlinear. This means that given two solutions, $g_{\m\n}^{\text{\tiny{(1)}}}$ and $g_{\m\n}^{\text{\tiny{(2)}}}$ then the sum of the two
\be
a g_{\m\n}^{\text{\tiny{(1)}}}+b g_{\m\n}^{\text{\tiny{(2)}}}
\ee
is not a solution. This is perhaps the main fact that prevents standard techniques (such as Fourier analysis) to be useful in this case. When we talk about linear equations, we are assuming superposition in this sense.

On the other hand, the  Fierz-Pauli (FP) equations \cite{Fierz} are linear equations that represent a consistent quantum field free theory of spin two $h_{\m\n}(x)$ in flat space $\eta_{\m\n}$ (although this can be generalized to an arbitrary spacetime background)
\be
\Box h_{\m\n}+\pd_\m\pd_\n h-\pd^\l\pd_\m  h_{\l\n}-\pd^\l\pd_\n  h_{\l\m} =0
\ee
where $h\equiv \eta^{\m\n} h_{\m\n}$. This theory can be proved to be the linear approximation to the Ricci tensor, id est, the first order in $\kappa\equiv \sqrt{8\pi G}$ in the expansion
\be
g_{\m\n}=\eta_{\m\n}+\kappa h_{\m\n}
\ee
we shall take units with $\kappa=1$ in the sequel, for simplicity.

In this paper we are going to exploit a technique to generate new Ricci flat spacetimes out of some other Ricci flat spacetime, which we will dub the {\em seed}. The idea is a generalization of the approach in \cite{Kerr} and originated in \cite{Gurses} and \cite{X}. 
\par
In the original work of Kerr and Schild  they examined  Ricci flat metrics for which there is a gauge of the form
\be
g_{\m\n}=\eta_{\m\n}+ l_\m l_\n
\ee
(where $\eta_{\m\n}$ is the flat Minkowski metric) and they found all of those  trough quite heavy calculations using a clever tretrad. In this paper  by using an apparently coarser approach we shall find unexpected relationships with the flat space spin 2 Fierz-Pauli equation.

\section{The generalized  Fierz-Pauli equation.}
In this section, we present the generalization of Fierz-Pauli equation for an arbitrary background. Using expansion of the metric with $\kappa=1$
\be \label{m1}g_{\m\n}=\bg_{\m\n}+ h_{\m\n}\ee
the first step in the computation is to calculate the inverse metric, $g^{\m\n}$. This is only possible as an infinite  power series; there is no closed formula in the general case. The same thing happens with the determinant of the metric. The one exception is precisely the hypothesis
\bea \label{h}&h_{\m\l}h^\l_\n=0\nonumber\\
&h=0
\eea
in which case  the inverse of the metric simply reads
\be
g^{\m\n}=\bg^{\m\n}-h^{\m\n}
\ee
and this expansion terminates; there are no more terms in it.

Next, let us compute an \textit{explicit} and \textit{exact} expressions of the Ricci tensor.  Cartan's connection (that is the Christoffel connection corresponding to the full metric $g_{\m\n}$) can be easily computed as
\be \Gamma^\m_{\a\b}=\bar{\Gamma}^\m_{\a\b}+C^\m_{\a\b}\ee
where
\be C^\m_{\a\b}=\frac{1}{2}\left[\bn_\a(h_{\b}^{~\m})+\bn_\b(h_{\a}^{~\m})-\bn^\m(h_{\a\b})+h^{\m\l}\bn_\l(h_{\a\b})\right]\ee
with the usefull property of  $C^\l_{\l\a}=0$
Please note that this expression is  also exact.

The corresponding  Ricci tensor of the metric \eqref{m1} is
\be R_{\m\n}=\bR_{\m\n}+\bn_{\r}C^\r_{\m\n}-C^\r_{\m\s}C^\s_{\n\r}\ee
After a little bit of algebra one gets
\be R_{\m\n}=\bR_{\m\n}+R_{\m\n}^{\text{\tiny{(1)}}}+R_{\m\n}^{\text{\tiny{(2+3)}}}\ee
where $R_{\m\n}^{\text{\tiny{(n)}}}$ is the expansion at order $n$ of the Ricci tensor. To be specific,
\bea
R_{\m\n}^{\text{\tiny{(1)}}}&=\frac{\kappa}{2}\left[\bn_\l\bn_\m(h_{\n}^{~\l})+\bn_\l\bn_\n(h_{\m}^{~\l})-\bar{\Box}(h_{\m\n})\right]\nonumber\\
R_{\m\n}^{\text{\tiny{(2+3)}}}&=\frac{\kappa^2}{2}\Big[\bn_\s\left[h^{\s\r}\bn_\r(h_{\m\n})\right]-\bn_\s (h_{\m}^{~\r})\bn_\r(h_{\n}^{~\s})+\nonumber\\
&+\bn^\r (h_{\m}^{~\s})\bn_\r(h_{\n\s})-\frac{1}{2}\bn_\m (h_{\r\s})\bn_\n(h^{\r\s})\Big]\nonumber\\
&=-\frac{\kappa^3}{2}h^{\r\s}\bn_\r (h_{\m\l})\bn_\s(h_{\n}^{~\l})\eea

To sum up,  the Ricci tensor of the metric \eqref{m1} with the condition \eqref{h} reduces to
\be \label{exact}R_{\m\n}=\bR_{\m\n}+R_{\m\n}^{\text{\tiny{(1)}}}+R_{\m\n}^{\text{\tiny{(2+3)}}}\ee
where the first term, which is linear in the deformation is exactly the linear Fierz- Pauli operator in the background $\overline{g}_{\m\n}$ with $h\equiv \overline{g}_{\m\n}h^{\m\n}=0$ and the second one  is nonlinear and more complicated and contains all the nonlinear terms up to third order in the deformation. We stress that in spite of the seemingly perturbative approach, this expansion also terminates; there are no more terms in it.

The surprising thing is that the terms of Ricci tensor obey
\be\label{teorema}
\{ R^{\text{\tiny{(1)}}}_{\a\b}=0\} \Longrightarrow \{R^{\text{\tiny{(2+3)}}}_{\a\b}=0\}
\ee
The proof is straightforward algebra. 
\par
To be specific, let us give details on the simplest case. Assume that the perturbation is a Kerr-Schild metric 
\be h_{\m\n}=l_\m l_\n\ee
where $l_\m$ is a nilpotent vector, $l^2=0$,  in  which  case \eqref{h} is trivially satisfied.  It has already been pointed out that in this case, and only in this case, the perturbative expansion has got only a finite number of terms.
\par
Throughout  this work we employ  the standard {\em optical variables}, namely, the {\em acceleration}
\be
\dot{l}^\a\equiv l^\l \bn_\l l^\a
\ee
(which vanishes if the vector field is geodesic) and the {\em expansion}
\be
\theta\equiv \bn_\l l^\l\ee
To be specific, the first order in the deformation reads 
\be
R^{\text{\tiny{(1)}}}_{\a\b}= {1\over 2}\left[\bn_\m \bn_\a ( l^\m l_\b )+\bn_\m \bn_\b (l^\m l_\a)-\bar{\Box} (l_\a l_\b)\right]
\ee
the remaining  terms in the expansion of the Ricci tensor are given by
\bea
& R^{\text{\tiny{(2+3)}}}_{\m\n}=\frac{1}{2}\Big\{\theta\left(\dot{l}_\m l_\n+l_\m\dot{l}_\n\right)+\ddot{l}_\m l_\n+\dot{l}_\m\dot{l}_\n+l_\m\ddot{l}_\n-\dot{l}^\r\bar{\nabla}_\r\left(l_\m l_\n\right)-\nonumber\\
&-l_\m l_\n \bn^\r l_\s\bn^\s l_\r+l_\m l_\n\bn^\s l^\r\bn_\s l_\r-l_\m l_\n \dot{l}_\r \dot{l}^\r\Big\}
\eea

Assuming 
\be R^{\text{\tiny{(1)}}}_{\m\n}=0\ee
and contracting with $l^\m l^\n$ yields 
\be
l^\m l^\l  l^\n \bn_\l \bn_\n l_\m=l^\l l^\n\bn_\l\left(l^\m\bn_\n l_\m\right)-l^\l l^\n\bn_\l l^\m \bn_\n l_\m=-\dot{l}_\m\dot{l}^\m=0
\ee
then, because $\dot{l}^\m l_\m=0$ there must exist  an scalar field $\phi$ such that
\be\label{fi}
\dot{l}^\a\equiv l^\l \bn_\l l^\a= \phi\, l^\a
\ee
As has already been pointed out, the  vector field is geodesic whenever $\phi=0$.
\par

Contracting $R^{\text{\tiny{(1)}}}_{\m\n}=0$ with $l^\m$  and using  this, we get
\bea
&l^\m l_\n \bar{\Box} l_\m -l^\m l_\n \bn_\l \bn_\m l^\l-l^\m\bn_\l l^\l \bn_\m l_\n-l^\m\bn_\m l^\l\bn_\l l_\n-l^\m l^\l \bn_\l \bn_\m l_\n -l^\m l^\l  \bn_\l \bn_\n l_\m=\nonumber\\
&=\left(\bn_\l l_\m\bn^\l l^\m +\dot{\theta}+\dot{\phi}+\phi^2+\theta\phi\right) l_\n=0
\eea
so that
\be\label{cuad}
\bn_\l l_\m\bn^\l l^\m =-(\dot{\theta}+\dot{\phi})-\phi(\phi+\theta)- \bR_{\a\b}l^\a l^\b
\ee
On the other hand computing $\dot{\theta} $ leads to
\be\label{teta}
\bn_\m l^\n \bn_\n l^\m=\dot{\phi}-\dot{\theta}+\phi\theta- \bR_{\a\b}l^\a l^\b
\ee
Collecting all the results
\be R^{\text{\tiny{(2+3)}}}_{\m\n}=0\ee

Then not only is the expansion of the Ricci tensor finite, but also the vanishing of the linear term implies the vanishing of the whole expansion. This is the more outstanding given that $R^{\text{\tiny{(1)}}}_{\a\b}$ is linear in the deformation $h_{\m\n}\equiv l_\m l_\n$, whereas $R^{\text{\tiny{(2+3)}}}_{\a\b}$ is quadratic and cubic in the deformation. This important theorem \eqref{teorema}  is in fact a simple consequence of  the previous equations \eqref{fi}\eqref{cuad}\eqref{teta}.
\par
We find it noteworthy that in this formalism solutions of the Fierz-Pauli equation in a Ricci flat background  automatically yield solutions of the vacuum Einstein's equations, that is new Ricci flat spacetimes.
\par
Next let us proceed to the study of some explicit examples of applications of our master formula \eqref{teorema}, starting with the simplest ones.
\subsection{Flat space as seed.}
Let us begin by considering flat spacetime as our background. In this case we generate the Kerr-Schild family of solutions \cite{Kerr}.

Assume for simplicity  a static radial ansatz
\be
l_\m=f(x^i)\left(1, -{x^j\over r}\right)
\ee
it turns out that the only function that works is  $f(r)=C r^{- 1/2}$  which  corresponds to a black hole  centered at the origin
\be
\text{d}s^2=\left(\eta_{\m\n}+l_\m l_\n\right)\text{d}x^\m \text{d}x^\n= \eta_{\m\n} \text{d}x^\m \text{d}x^\n-{r_s\over r}(\text{d}t+\text{d}r)^2
\ee
deformed with 
\be l_{\m}\equiv \sqrt{\frac{r_s}{r}}\left(1,-\frac{x^j}{r}\right)\ee

It can be easily checked that it obeys $R^{\text{\tiny{(1)}}}_{\a\b}=0$. In that way we can easily reach all spacetimes of the Kerr-Schild family \cite{Stephani}\cite{Kerr}\cite{Adamo}.

\subsection{Ricci flat space as seed.}

Consider now a more complicated background, namely  a non-flat but still Ricci flat spacetime. 
\begin{itemize}
	\item We shall choose the Kasner solution for simplicity. The metric reads
	\be
	\text{d}s^2= \text{d}t^2-\sum_{i=1}^3  t^{2 p_i} \text{d}x_i^2
	\ee
	through some ponderous trial and error, we found a solution of FP in a Kasner background which is given by
	\be
	l_\m=\left(1,t^{p_1},0,0\right)
	\ee
	then
	\be
	\text{d}s^2= 2 \text{d}t^2+2 t^{p_1}\text{d}t \text{d}x-t^{2p_2} \text{d}y^2-t^{2p_3} \text{d}x^2
	\ee
	should be a new Ricci flat metric. It is in fact so.  The same thing happens with the related  vectors
	\be
	l_\m=\left(1,0,t^{p_2},0\right)
	\ee
	yielding
	\be
	\text{d}s^2= 2 \text{d}t^2-t^{2p_1} \text{d}x^2+2 t^{p_2} \text{d}t \text{d}y-t^{2p_3} \text{d}z^2
	\ee
	and finally
	\be
	l_\m=\left(1,0,0,t^{p_3}\right)
	\ee
	with
	\be\label {z}
	\text{d}s^2= 2 \text{d}t^2-t^{2p_1} \text{d}x^2-t^{2p_2} \text{d}y^2+2t^{p_2} \text{d}t \text{d}z
	\ee
	We were able to generate new Ricci flat spacetimes out of a Ricci flat seed.  All those spacetimes are closely related to the Kasner family. Consider the last case as an example. It lacks the isometry
	\be
	z\rightarrow -z
	\ee
	
	In fact the spacetime metric can be written as 
	\be
	\text{d}s^2= 2 \left(\text{d}t+{1\over 2} t^{p_2} dz\right)^2-t^{2p_1} \text{d}x^2-t^{2p_2} \text{d}y^2-{1\over 2} t^{2 p_2} dz^2
	\ee
	\item 
	Another possibility is the plane wave, in Brinkmann coordinates. The background metric \cite{Garriga} reads in this case
	\be
	\text{d}s^2=\text{d}u\text{d}v-H_{ab}[u]x^ax^b\text{d}u^2-\d_{ij}\text{d}x^i\text{d}x^j
	\ee
	this metric has an  only nonvanishing component of the Ricci tensor 
	\be
	R_{uu}=-\d^{ab}H_{ab}[u]\equiv -H^a_a[u]
	\ee
	obviously the manifold is Ricci flat whenever the transverse matrix $H_{ab}$ is traceless.
	
	Consider the constant null vector with associated one form
	\be
	l_\m \text{d}x^\m= \sqrt{v} \text{d}u
	\ee
	it is straightforward to check that
	\bea
	&R^{\text{\tiny{(1)}}}_{\a\b}=R^{\text{\tiny{(2+3)}}}_{\a\b}=0\eea
	the deformed metric reads
	\be
	\text{d}s^2=\text{d}u\text{d}v-H_{ab}[u]x^ax^b\text{d}u^2-\d_{ij}\text{d}x^i\text{d}x^j+vdu^2
	\ee
	which is Ricci flat, when $H^a_a[u]=0$. Those spacetimes again represent gravitational waves of a sort.
\end{itemize}
To sum up, in this section we have found that the vanishing of the first term in the expression for the Ricci tensor, $R^{\text{\tiny{(1)}}}_{\a\b}=0$ (Fierz-Pauli) implies the vanishing of the second one $R^{\text{\tiny{(2+3)}}}_{\a\b}=0$. We have applied this to some simple examples, finding several new solutions.
\section{Sources}
It is also interesting to test this approach when there is a source in Einstein's equations. We are only able to do this in some simple cases. Consider for example a constant curvature spacetime, like de Sitter or anti de Sitter (the sign of the cosmological constant is not material in the sequel)

\be \text{d}s^2_{\text{\tiny{(CCS)}}}=\bg_{\m\n}\text{d}x^\m\text{d}x^\n=\text{d}t^2- e^{2Ht}\d_{ij}\text{d}x^i\text{d}y^j\ee
where $H$ is a constant. The Ricci tensor reads
\be \bR_{\m\n}=-3H^2 \bg_{\m\n}\ee
and obeys Einstein's equations sourced by an energy-momentum tensor
\be \bR_{\m\n}=\kappa^2\left(T_{\m\n}-\frac{1}{2}\bg_{\m\n}T\right)\ee
with
\be T_{\m\n}=\frac{3H^2}{\kappa^2}\bg_{\m\n}\ee
corresponding to a vacuum energy density
\be \r=-p=\frac{3H^2}{\kappa^2}\ee

Consider  a deformation of the same constant curvature de Sitter background with a null vector field which is isotropic in the spatial directions
\be l_\m=\left(1,\frac{1}{\sqrt{3}}e^{Ht},\frac{1}{\sqrt{3}}e^{Ht},\frac{1}{\sqrt{3}}e^{Ht}\right)\ee
The resulting spacetime metric looks quite involved
\bea
&ds^2= 2\,  dt^2-{2\over 3} e^{2 H t}\left( dx^2+ dy^2+ dz^2\right)+{2\over \sqrt{3}}e^{Ht}\left( dt dx + dt dy+ dt dz\right)+\nonumber\\
&+{2\over 3} e^{ 2 H t} \left( dx dy+ dx dz+ dy dz\right)
\eea
In spite of the appearances this cosmological metric shares rotational invariances with the seed spacetime.

\par
The perturbation $h_{\m\n}$ now  induces a nonvanishing $R^{\text{\tiny{(1)}}}_{\a\b}$ tensor, namely
\be R^{\text{\tiny{(1)}}}_{\m\n}=H^2\begin{pmatrix}
	0 & -\sqrt{3}e^{Ht} & -\sqrt{3}e^{Ht}&-\sqrt{3}e^{Ht}\\
	-\sqrt{3}e^{Ht} & -4e^{2Ht} & -e^{2Ht}&-e^{2Ht}\\
	-\sqrt{3}e^{Ht} & -e^{2Ht} & -4e^{2Ht}&-e^{2Ht}\\
	-\sqrt{3}e^{Ht}&-e^{2Ht}&-e^{2Ht}&-4e^{2Ht}
\end{pmatrix}
\ee
and
\be R^{\text{\tiny{(2+3)}}}_{\m\n}=H^2\begin{pmatrix}
	3 &\sqrt{3}e^{Ht} & \sqrt{3}e^{Ht}&\sqrt{3}e^{Ht}\\
	\sqrt{3}e^{Ht}& e^{2Ht} & e^{2Ht}&e^{2Ht}\\
	\sqrt{3}e^{Ht} & e^{2Ht} & e^{2Ht}&e^{2Ht}\\
	\sqrt{3}e^{Ht}&e^{2Ht}&e^{2Ht}&e^{2Ht}
\end{pmatrix}
\ee
Now there is again a conspiracy between the Ricci tensor of the background and the $R^{\text{\tiny{(1)}}}_{\m\n}$ and $ R^{\text{\tiny{(2+3)}}}_{\m\n}$ tensors  to end up again with a Ricci flat spacetime.
\bea R_{\m\n}&=0\eea

To summarize what we have done so far, we showed that  beginning with a background which is not a vacuum solution of Einstein field equations (it needs a constant vacuum energy density as a source) we end up with a new vacuum solution.
\par
Let us finish this paragraph with a comment on the vanishing of the counterterms. in General Relativity.
The first quantum correction of General Relativity which does not vanish on shell is well-known, \cite{tHooft} and \cite{Goroff:1985th} to be the two-loop one
\bea\label{count}
\Delta \mathcal{L}&=\frac{\sqrt{|g|}}{\e}\left(\frac{7}{10}\bR_{\m\n}\bR^{\m\n}+\frac{1}{60}\bR^2+\frac{209}{2880}\bR_{\m\n\a\b}\bR^{\a\b\l\t}\bR_{\l\t}^{~~\m\n}\right)\eea
Let is now  assume a particular  background metric such as  
\be g_{\m\n}=\bg_{\m\n}+\kappa n_{\m\n}+\kappa h_{\m\n}\ee 
with $n_{\m\l}\overline{g} ^{\l\s}n_{\n\s}=0$ and $n\equiv \overline{g}^{\m\n} n_{\m\n}=0$

 Goroff-Sagnotti's operator  can be computed  to the lowest order in $\kappa$ after a rather long calculation
\bea\label{k3}
&\bR_{\m\n\a\b}\bR^{\a\b\l\t}\bR_{\l\t}^{~~\m\n}=\kappa^3\bn^\r\bn^\s n^{\a\b}\Big\{3\left(\bn_\s\bn_\m n_{\b\n}-\bn_\b\bn_\m n_{\s\n}\right)\bn^\n\bn_\r h_\a^\m+\nonumber\\
&+\bn_\a\bn^\n n_{\s\m}\bn^\m\bn_\b n_{\r\n}-\bn_\s\bn^\n n_{\a\m}\bn_\n\bn_\r n_\b^\m\Big\}+\mathcal{O}(\kappa^4)\eea

It is plain that  in a {\em vanishing scalar invariant}  (VSI) spacetime all   polynomial curvature invariants  vanish so that the same thing happens with the GS counterterm.

We can slightly generalize this result. It can be proved by explicit computation that  the metric (which is not VSI)
\be g_{\m\n}=\text{d}u\text{d}v-\d_{ij}\text{d}x^i\text{d}x^j- H[u,x,y]\d_{\m u}\d_{\n u}+\kappa h_{\m\n}\ee
has the one-loop and two-loops counterterm zero.

\section{Is there superposition?}
It would seem that we have somewhat magically, traded a nonlinear equation (Ricci flatness) by a linear one (Fierz-Pauli) in which case it seems that there must be superposition of solutions. Things are more complicated though.
\par
First of all, in order to find the first new Ricci flat metric,  the null vector has to be a solution of the FP equation
\be
R^{\text{\tiny{(1)}}}_{\a\b}=0
\ee
if now we want to consider the defomed spacetime
\be
g_{\m\n}\equiv \bg_{\m\n}+ l_\m l_\n
\ee
as a new seed, we need to solve
\be
R_{\m\n}=\bR_{\m\n}
\ee
which is different equation altogether. There is no question of superposition. 
\par

But there is more. 
\par

Even if we consider the same FP equation,  given two different  nilpotent solutions of the   equation $R^{\text{\tiny{(1)}}}_{\a\b}=0$ , say
\be
h_{\m\n}\equiv l^{\text{\tiny{(1)}}}_\m  l^{\text{\tiny{(1)}}}_\n
\ee
and 
\be
h_{\m\n}\equiv l^{\text{\tiny{(2)}}}_\m  l^{\text{\tiny{(2)}}}_\n
\ee
their sum
\be
h_{\m\n}\equiv l^{\text{\tiny{(1)}}}_\m  l^{\text{\tiny{(1)}}}_\n+  l^{\text{\tiny{(2)}}}_\m  l^{\text{\tiny{(2)}}}_\n
\ee
is of course again a solution of the Fierz-Pauli equations $R^{\text{\tiny{(1)}}}_{\a\b}=0$, {\em but not necessarily nilpotent}. In order to fulfill this condition, it is necessary that there exists a third null vector, $l^{\text{\tiny{(3)}}}_\m $ such that
\be
h_{\m\n}=l^{\text{\tiny{(3)}}}_\m  l^{\text{\tiny{(3)}}}_\n
\ee

This happens only when
\be
l_{\text{\tiny{(1)}}}^\l l^{\text{\tiny{(2)}}}_\l=0
\ee
which in turn enforces
\be
l_\l^{\text{\tiny{(1)}}}=  l_\l^{\text{\tiny{(2)}}}=l_\l^{\text{\tiny{(3)}}}
\ee
\par
Consider for example  two black hole located at different points where
\be
h_{\m\n}=
\begin{pmatrix}{1\over r}+{1\over |\vx-\va|}&-{x^j\over r^2}-{x^j-a^j\over|\vx-\va|^2} \\-{x^j\over r^2}-{x^j-a^j\over|\vx-\va|^2}&{ x^i x^j\over r^3}+{(x^i-a^i)(x^j-a^j) \over  |\vx-\va|^3}\end{pmatrix}
\ee
it is clear 
\be \det\,h_{\m\n}\neq 0\ee
and the  deformed spacetime  is not Ricci flat.

This means that in spite of the fact that $R^{\text{\tiny{(1)}}}_{\a\b}=0$  is a linear equation (the crucial point is that it depends on the background), superposition in the present context is still an exceptional property of Ricci flat spacetimes. It basically reduces to scaling \cite{Alvarez}.
\par
To be specific, consider a spherical black hole (BH) in Eddington coordinates
\bea \label{KS} &\text{d}s^2_{\text{\tiny{(1)}}}=\eta_{\m\n} \text{d}x^\m \text{d}x^\n-{r_s^{\text{\tiny{(1)}}}\over r}(\text{d}t+\text{d}r)^2\eea
this metric supports a light vector
\be l_{\m}\equiv \sqrt{\frac{r_s^{\text{\tiny{(1)}}}}{r}}\left(1,-\frac{x^j}{r}\right)\ee
it can be easily verified that this metric is Ricci flat and unimodular.

\par
Let us now consider a superposition of two such metrics
\bea \text{d}s^2_{\text{\tiny{(1+2)}}}&=\eta_{\m\n} \text{d}x^\m \text{d}x^\n-{r_s^{\text{\tiny{(1)}}}+r_s^{\text{\tiny{(2)}}}\over \sqrt{2}r}(\text{d}t+\text{d}r)^2\eea
it is plain that this superposition is Ricci flat as well.
\par

The Schwarzschild radius of the superposition is
\be
r_s^{(12)}={r_s^{(1)}+r_s^{(2)}\over \sqrt{2}}
\ee
This of course, the same relationship of the mass of the combined spacetime containing two black holes on top of each other with the parent masses. It is to be remarked that this mass is {\em smaller} than the sum of the two initial masses.
The entropy, in turn, is
\be
S_{(12)}=\pi \left(r_s^{ (12)}\right)^2=\pi {\left(r_s^{(1)}+r_s^{(2)}\right)^2\over 2}=S_{(1)}+S_{(2)} -{\pi\over 2} \left(r_s^{(1)}-r_s^{(2)}\right)^2
\ee
which again, {\em smaller} than the sum of the two parent entropies. We do not know any physical interpretation of this fact. Were we to add $n$ copies of the BH, a similar rescaling leas to 
\be
r_s^{(n)}={1\over \sqrt{n}}\sum_{i=1}^n r_s^{(i)}
\ee
again,  {\em smaller} than the sum of the two initial masses.
The entropy reads
\be
S_{(n)}=\pi \left(r_s^{ (n)}\right)^2=\sum_{i=1}^nS_{(i)}-\pi {n-1\over n}\sum_{i=1}^n (r_s^{(i)})^2+{\pi\over n} \prod_{i\neq j} r_s^{(i)}r_s^{(j)}
\ee

To summarize, in this section we have showed that in spite of the central r\^ole that the linear Fierz-Pauli equation plays in our approach, there is no superposition of solutions in general. Only a given spacetime can be put on top of itself, as has been explicitly done with black holes; this is essentially scaling.

\section{Conclusions}
In this paper we have studied nilpotent deformations of an arbitrary background (or seed), {\em id est},
\be
g_{\m\n}(x)=\bg_{\m\n}(x)+l_\m l_\n
\ee
with $l^2=0$. Those are an extension of the original Kerr-Schild ansatz, which obviously covers the full space of metrics, just by starting with an appropiate seed metric.
\par
A general formula for the ensuing Ricci tensor has been obtained, namely.
\be
R_{\m\n}=\bR_{\m\n}+ R^{\text{\tiny{(1)}}}_{\m\n}+R^{\text{\tiny{(2+3)}}}_{\m\n}
\ee
where $R^{\text{\tiny{(1)}}}_{\m\n}$ is a linear operator to wit, the Fierz-Pauli one and $R^{\text{\tiny{(2+3)}}}_{\m\n}$ is a non-linear operator. They have the remarkable property that
\be
\{ R^{\text{\tiny{(1)}}}_{\a\b}=0\} \Longrightarrow \{R^{\text{\tiny{(2+3)}}}_{\a\b}=0\}
\ee
\par
When the seed is taken as flat space, we recover the Kerr-Schild family of Ricci flat solutions.
\par
We have also considered nontrivial Ricci flat spaces as seed, such as Kasner spacetime or plane waves, and also Ricci non flat seeds, concretely de Sitter space, which solves Einstein's equations with a source. It is plain that we have only begun to scratch the surface of the possibilities of our theorem to find new vacuum solutions of Einstein's equations. 
\par
A grand view of the situation is as follows. 
\par
The space of all Ricci flat metrics in a given spacetime is divided into {\em orbits}, in such a way that it is not possible to travel from one orbit to a different one through the deformations used in this paper.
\par
For example, the orbit generated by flat space comprises all Kerr-Schild spacetimes, but it is known that constant curvature spaces do not belong to this orbit. This shows that there are at least two different orbits.
\par
We have finally pointed out that in spite of the fact that in some sense, for this particular family of spacetimes, the Ricci flat condition has beed traded by the Fierz-Pauli equation, which is linear, there is no superposition of spacetimes in general. Only scaling survives \cite{Alvarez} in which some spacetime is put on top of itself.

\section{Acknowledgements.}

One of us (EA) is grateful for  stimulating discussions in HKUST with Andy Cohen and Yi Wang.  We acknowledge partial financial support by the Spanish MINECO through the Centro de excelencia Severo Ochoa Program  under Grant CEX2020-001007-S  funded by MCIN/AEI/10.13039/501100011033.
We also acknowledge partial financial support by the Spanish Research Agency (Agencia Estatal de Investigaci\'on) through the grant PID2022-137127NB-I00 funded by MCIN/AEI/10.13039/501100011033/ FEDER, UE.
All authors acknowledge the European Union's Horizon 2020 research and innovation programme under the Marie Sklodowska-Curie grant agreement No 860881-HIDDeN and also byGrant PID2019-108892RB-I00 funded by MCIN/AEI/ 10.13039/501100011033 and by ``ERDF A way of making Europe''.

\newpage

\newpage
  
\end{document}